\documentclass[a4paper,twocolumn,11pt,accepted=2022-06-16]{quantumarticle}
\pdfoutput=1
\usepackage[utf8]{inputenc}
\usepackage[english]{babel}
\usepackage[T1]{fontenc}
\usepackage{amsmath,mathtools,amsfonts,amssymb}
\usepackage{hyperref}

\usepackage{tikz}
\usepackage{lipsum}
\usepackage{braket}
\usepackage{subfigure}
\graphicspath{ {./figures/} }
\usepackage[numbers,sort&compress]{natbib}

\begin{document}

\title{Quantum scattering as a work source}

\author{Samuel L. Jacob}
\email{samuel.lourenco@uni.lu}
\affiliation{Complex Systems and Statistical Mechanics, Physics and Materials Science Research Unit, University of Luxembourg, L-1511 Luxembourg, G.D. Luxembourg}
\affiliation{Kavli Institute for Theoretical Physics, University of California, Santa Barbara, CA 93106 Santa Barbara,  U.S.A.}

\author{Massimiliano Esposito}
\email{massimiliano.esposito@uni.lu}
\affiliation{Complex Systems and Statistical Mechanics, Physics and Materials Science Research Unit, University of Luxembourg, L-1511 Luxembourg, G.D. Luxembourg}
\affiliation{Kavli Institute for Theoretical Physics, University of California, Santa Barbara, CA 93106 Santa Barbara,  U.S.A.}

\author{Juan M. R. Parrondo}
\email{parrondo@fis.ucm.es}
\affiliation{Departamento de Estructura de la Materia, F\'isica T\'ermica y Electr\'onica and  GISC, Universidad Complutense Madrid, 28040 Madrid, Spain}

\author{Felipe Barra}
\email{fbarra@dfi.uchile.cl }
\affiliation{Departamento de F\'isica, Facultad de Ciencias F\'isicas y Matem\'aticas, Universidad de Chile, 837.0415 Santiago, Chile}
\affiliation{Kavli Institute for Theoretical Physics, University of California, Santa Barbara, CA 93106 Santa Barbara,  U.S.A.}


\maketitle


\begin{abstract}
We consider a collision between a moving particle and a fixed system, each having internal degrees of freedom. We identify the regime where the motion of the particle acts as a work source for the joint internal system, leading to energy changes which preserve the entropy. This regime arises when the particle has high kinetic energy and its quantum state of motion is broad in momentum and narrow in space, whether pure or mixed.
In this case, the scattering map ruling the dynamics of the internal degrees of freedom becomes unitary and equivalent to that of a time-dependent interaction between the internal degrees of freedom of the colliding systems. It follows that the kinetic energy lost by the particle during the autonomous quantum collision coincides with the work performed by the time-dependent interaction.
Recently, collisions with particles were shown to act as heat sources under suitable conditions; here we show that they can also act as work sources. This opens interesting perspectives for quantum thermodynamics formulations within scattering theory.

\end{abstract}





\section{Introduction}

The dynamics of quantum systems interacting with other systems is in general described by completely positive and trace preserving (CPTP) maps~\cite{Breuer2007,Rivas2012}. However, there may be regimes in which this map acts approximately as a unitary transformation and thus the system's von Neumann entropy remains constant. One example is the semi-classical regime of light-matter interaction, where the driving of an atom by the electromagnetic field is modelled by a time-dependent interaction, leading to a unitary evolution for the atom \cite{Jaynes1963,Cohen1991vol2,Cohen1998,Haroche2006}. Another example is when a particle travelling semi-classically is used to measure the time associated to a quantum process \cite{Aharonov1961,Woods2019}. Identifying these unitary regimes is crucial not only from a dynamical point of view, but also for quantum thermodynamics. Indeed, although the thermodynamic notion of work for quantum systems is still debated \cite{Hanggi2016,Niedenzu2019}, the energy changes induced by these unitary evolutions can often be interpreted as work \cite{Esposito2010,Kosloff2013,Acin2015,Anders2016,Binder2018}.

Scattering theory plays a central role in quantum physics, from high energy physics to mesoscopic physics \cite{Razavy2003,Belkic2004,Taylor2006,Nazarov2009,vonOppen2018} and open quantum system theory \cite{Dumcke1985,Zeilinger2003,Hornberger2006,Hornberger2007,Hornberger2008,Hornberger2008}. It also provides the most direct connection between quantum theory and experimental observables. 
The fundamental object of the theory -- the scattering operator -- is a unitary energy preserving transformation, similar to those considered in the resource theory of quantum thermodynamics~\cite{Lostaglio2019}. Considering maps generated by collision events may thus be used to bring resource theory closer to experiments. 

In a recent work~\cite{Barra2021}, we studied the collision between a fixed system and a travelling particle, and showed that the dynamics of the (joint) internal degrees of freedom is ruled by a CPTP map which, under certain conditions, induces decoherence and thermalization. 

In the present paper, we move in the opposite direction and show that the joint internal dynamics becomes unitary when 
the particle has a high kinetic energy and its motion is described by a state that is squeezed in position and broad in momentum. Our results extend the textbook semi-classical treatment of a point particle scattered by an external potential \cite{Landau1977,Sakurai1993} to situations where the particle has an internal structure.

We furthermore show that, in this regime, the joint internal dynamics, as well as its thermodynamics, is equivalent to that of a unitary dynamics generated by turning on and off a time-dependent interaction between the system and particle's internal degrees of freedom. 
Such time-dependent models of collisions have been used in recent years to study thermodynamics of quantum systems based on repeated interactions \cite{Barra2015,Strasberg2017,Rodrigues2019,Guarnieri2020}. The time-dependent work in these models is here shown to arise more fundamentally within scattering theory from the kinetic energy changes of the colliding particle.

Our results are relevant for quantum thermodynamics in general ~\cite{Anders2016,Binder2018} and for developing autonomous work extraction devices in particular~\cite{Monsel2020}. Together with the results in Ref.~\cite{Barra2021}, they form the basis for a comprehensive approach towards the design of thermodynamic processes using scattering.
 
The rest of the paper is organized as follows. In section \ref{secII} we introduce the scattering map and the time-dependent model, and state our results. In section \ref{sec:td}, we sketch the proof of the results for the scattering map and prove it for the time-dependent model. The technical parts, such as the semi-classical treatment of the collision, are detailed in the Appendices. Section \ref{secV} illustrates our findings using an example and conclusions are drawn in section \ref{secVI}. 
 

\section{The Setup and Results}
\label{secII}

We start by considering two quantum systems $A$ and $B$. The joint system $Y$ has Hamiltonian $H_{Y} = H_A \otimes \mathbb{I}_B + \mathbb{I}_A \otimes H_B$, where  $\mathbb{I}_A$ denotes the identity operator on the Hilbert space of $A$ (equivalently for $B$).
The spectrum $H_{Y} \ket{j}=e_j\ket{j}$ of $H_Y$ is finite, $\{ \ket{j} \}_{j=1}^{N}$ and $e_j$ increases with $j$. We define $\Delta_{j'j} \equiv e_{j'}-e_{j}$ and the characteristic energy scale $\Delta_Y \equiv  \Delta_{N1}$.

\subsection{Scattering map}

\begin{figure*}[t!]
\centering
\includegraphics[width=0.7\textwidth]{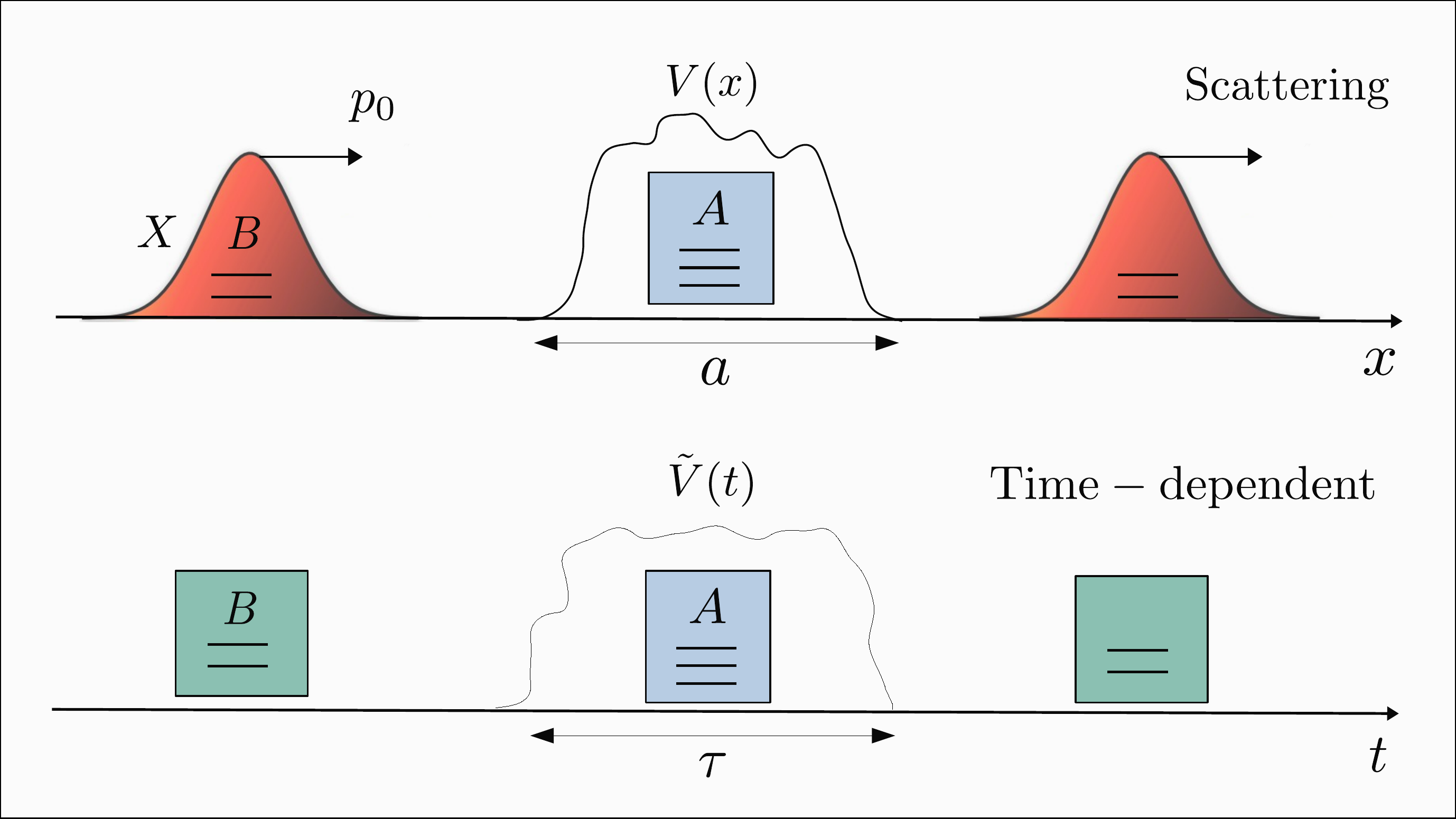}
\caption{\label{fig:models} Scattering and time-dependent setups considered in this study. In the former, the interaction happens autonomously in space through a potential $V(x)$: system $A$ is fixed while the incoming particle has kinetic degrees of freedom $X$ and internal degrees of freedom $B$. In the latter, the interaction between $A$ and $B$ happens in time through a time-dependent interaction $\tilde{V}(t)$. Note that the potentials $V(x)$ and $\tilde{V}(t)$ are generally not the same.}
\end{figure*}

We start by modelling the interaction between $A$ and $B$ by a collision. In a reference frame co-moving with the center of mass, only the reduced mass plays a role, but we simplify the treatment by fixing system $A$ and consider a collision with a particle of mass $m$ moving in one dimension with internal structure $B$. The kinetic degree of freedom of the particle is denoted by $X$ (see Fig.~\ref{fig:models}).
The Hamiltonian of the full system reads $H = H_0 + \mathcal{V}(x)$, where $H_0 = H_Y \otimes \mathbb{I}_X + \mathbb{I}_Y \otimes p^2/2m$. The kinetic energy operator accounts for the motion of the particle.
The interaction between the particles is described by the operator $\mathcal{V}(x) = \nu \otimes V(x)$, where $V(x)$ is a non-vanishing function only inside the interval $x \in (-a/2,a/2)$ and $\nu$ is the interaction on $Y$. We take the full system to be initially in a factorized state $\rho_A \otimes \rho_B \otimes \rho_X$ with $\rho_X$ the state describing the kinetic degree of freedom of the particle. For example, for a pure state $\rho_X = \ket{\phi}\bra{\phi}$, where $\ket{\phi}$ is a wave packet, we have the average momentum $p_0=\bra{\phi}p\ket{\phi}$ and position $x_0=\bra{\phi}x\ket{\phi}$ with corresponding variances $\sigma_p^2$ and $\sigma_x^2$.
Scattering theory allows us to compute the final state of $Y$ after the collision
	\begin{align}
	\label{scattmap}
	\rho' =\mathrm{Tr}_{X} \big[ S \big( \rho_A \otimes \rho_{B} \otimes \rho_{X} \big) S^{\dagger} \big] \; ,
	\end{align}
where $\mathrm{Tr}_{X}$ denotes the partial trace over $X$ and $S$ is the unitary scattering operator~\cite{Belkic2004,Taylor2006}
\begin{align}
S=\lim_{t\to+\infty} e^{i\frac{t}{\hbar}H_0}e^{-i\frac{2t}{\hbar}H}e^{i\frac{t}{\hbar}H_0} \; .
\end{align}
Eq.~\eqref{scattmap} defines the scattering map ruling the state change of the internal system $Y$.
The change produced by $S$ reflects the full effect of the collision on the system's state without introducing an \textit{ad hoc} interaction time. 
Importantly, the scattering operator satisfies the commutation relation $[S,H_0]=0$ expressing total energy conservation (kinetic plus internal) in a collision between fixed system and particle.
The energy change in $Y$ is given by
	\begin{align}
	\label{kin}
	\Delta E = \mathrm{Tr}_Y \big[ H_{Y}  (\rho' - \rho_A \otimes \rho_B) \big] = -\Delta E_p \; ,
	\end{align} 
where $\Delta E_p = \mathrm{Tr}_X[(p^2 / 2m) (\rho'_X - \rho_X)]$ is the change in kinetic energy and $\rho'_X$ is the final state of the particle's motion, obtained by tracing over $Y$ instead of $X$ in Eq.~\eqref{scattmap}. The second equality in Eq.~\eqref{kin} follows from 
$[S,H_0]=0$. In general, the dynamics for the system $Y$ described by Eq.~\eqref{scattmap} is not unitary  and thus the associated entropy change for a collision $\Delta \mathcal{S} =\mathcal{S}(\rho')-\mathcal{S}(\rho_A\otimes\rho_B)$ is, in general, non-zero, preventing the identification of the energy change in Eq.~\eqref{kin} with work.

\subsection{Time-dependent model}

Time-dependent models describe the interaction of $A$ and $B$ for a time $\tau$ with Hamiltonian $H(t) =H_{Y} + V(t)$ (Fig.~\ref{fig:models}). The time-dependent interaction is given by $V(t) = \tilde{V}(t)\nu$, where $\tilde{V}(t)$ is a non-vanishing function only in the interval $t \in (-\tau/2,\tau/2)$ and $\nu$ is a time-independent operator. 
The evolution generated by the time-dependent model is unitary and, 
in the interaction picture, the density operator describing the state $Y$ after the interaction reads
	\begin{align}
	\label{tdmap}
	\rho_{\tau} & =  U_I(\tau) (\rho_A \otimes \rho_B) U^{\dagger}_{I}(\tau) \; ,
	\end{align}
where $U_I(\tau)$ is the unitary evolution operator in the interaction picture and the initial state is assumed factorized.
The energy change during the interaction is
    \begin{align}
	\label{work}
	W = \mathrm{Tr}[H_Y(\rho_{\tau} - \rho_A \otimes \rho_B)] \; ,
	\end{align}
and vanishes if $H_Y$ and $V(t)$ commute for all times. This energy change is interpreted as work~\cite{Esposito2010,Barra2015,Strasberg2017}
since the system is isolated and the von Neumann entropy $\mathcal{S}(\rho) = - k_B \mathrm{Tr}[\rho \log \rho] \geq 0$ is constant for a unitary time evolution $\Delta \mathcal{S}_{\tau} \equiv \mathcal{S}(\rho_{\tau}) - \mathcal{S}(\rho_A \otimes \rho_B)=0$. 

\subsection{Results}

We now state our first result. Under conditions to be specified below, the scattering map in Eq.~\eqref{scattmap} becomes the unitary transformation
	\begin{align}
	\label{result}
	\rho' = e^{-i \tau_{p_0} V/\hbar}~ (\rho_A \otimes \rho_B)~e^{i \tau_{p_0} V/\hbar} \; ,
	\end{align}
where the interaction time is $\tau_{p_0} \equiv ma/p_0$ and $V \equiv \langle V \rangle  \nu$ represents an effective interaction with
\begin{align}
\langle V \rangle \equiv \frac{1}{a}\int_{-\frac{a}{2}}^{\frac{a}{2}} V(x) dx \; . 
\end{align}
The time-dependent dynamics of Eq.~\eqref{tdmap} is also equivalent to Eq.~\eqref{result} after the substitution $\tau \rightarrow \tau_{p_0}$ and $\rho_{\tau} \rightarrow \rho'$, with the effective interaction being time-independent and given by $V \equiv \langle \tilde{V} \rangle  \nu$ where
\begin{align}
\langle \tilde{V} \rangle \equiv \frac{1}{\tau} \int_{-\frac{\tau}{2}}^{\frac{\tau}{2}} \tilde{V}(t) dt \; .
\label{avVt}
\end{align}
When the effective potentials and the interaction times are the same in the scattering map and in the time-dependent model, i.e. $\langle \tilde{V} \rangle = \langle V \rangle$ and $\tau = \tau_{p_0}$, the change of state of system $Y$ becomes the same in both cases.

Our second result follows directly from Eq.~\eqref{result} applied to Eq. \eqref{kin}: The energy change in $Y$ due to the collision, given by minus the change in kinetic energy of the particle, is equivalent to work since it occurs with a vanishing entropy change. Moreover, when $\langle \tilde{V} \rangle = \langle V \rangle$ and $\tau = \tau_{p_0}$ it equals to that of the time-dependent model in Eq. \eqref{work}.

\subsubsection*{Conditions of validity}

The conditions under which Eq.~\eqref{result} holds are:
\begin{itemize}
    \item[1)] The interaction time is much smaller than the time associated with the free evolution of the internal system $\tau,\tau_{p_0} \ll \hbar / \Delta_Y$.
    \item[2)]  The particle travels semi-classically over the potential $p_0^2/2m \gg V(x)$ and $p_0 a_{\rm min} \gg \hbar$. Here, $a_{\rm min}$ is the minimal length over which $V(x)$ varies.
    \item[3)] The particle's state of motion $\rho_X$ is fast, narrow in position and broad in momentum with respect to the internal system, as expressed by the inequalities $p_0 \gg \sigma_p \geq \hbar / 2 \sigma_x \gg m \Delta_Y /p_0$. 
\end{itemize}

Conditions 1 and 2 allows us to simplify the scattering map in Eq.~\eqref{scattmap}, while condition 3 is essential to preserve the coherence of the collision. If the last is not satisfied, system $Y$ decoheres \cite{Barra2021}. Importantly, condition 3 is valid for mixed states with arbitrary $\rho_X$, provided the average momentum, position, and corresponding variances are well defined. For a 
minimal uncertainty state (pure Gaussian state), we have $\sigma_p \sigma_x = \hbar / 2$ and condition 3 simplifies to $p_0 \gg \sigma_p \gg m \Delta_Y /p_0$.

Together, these conditions define a regime of high kinetic energies where the entanglement of the internal system $Y$ with the kinetic degree of freedom $X$ due to scattering is negligible. They are sufficient to achieve the unitary dynamics of $Y$ presented in Eq.~\eqref{result}, where the effective interaction time $\tau_{p_0}$ emerges from the scattering map Eq.~\eqref{scattmap} at high kinetic energies.

\section{Derivation} 
\label{sec:td}

\subsection{Scattering map}

We first discuss how to derive our results starting from Eq.~\eqref{scattmap}. We recall that
$H_0=H_Y+p^2/2m$ with $H_Y\ket{j}=e_j\ket{j}$ and $(p^2/2m) \ket{p} = E_p \ket{p}$. Here, $\{ \ket{p} \}$ are improper (non-normalizable) eigenstates whose position representation are plane waves $\braket{x|p} = \exp( \mathrm{i} p x / \hbar) / \sqrt{2 \pi \hbar}$ and $E_p = p^2 / 2m \geq 0$ is the kinetic energy. 
Due to the conservation of energy, the scattering operator in the eigenbasis of $H_{0}$, denoted by $\ket{p,j}\equiv\ket{p}\otimes \ket{j}$, is given by \cite{Taylor2006,Belkic2004,Barra2021}
    \begin{align}
	\label{expresionS}
	\braket{p',j'|S|p,j}& =\frac{\sqrt{|pp'|}}{m}\delta(E_{p}-E_{p'}-\Delta_{j'j}) \nonumber \\
	& \times s_{j'j}^{(\alpha' \alpha)}(E) \; ,
    \end{align}
where $s_{j'j}^{(\alpha' \alpha)}(E)$ is the scattering matrix at total energy $E = E_p + e_j$ and $\alpha={\rm sign}(p)$ and $\alpha'={\rm sign}(p')$ accounts for the initial and final direction of the momenta,
which can be positive or negative. The pairs $(++),(+-),(-+),(--)$ correspond to transmission from the left, reflection from the left, reflection from the right and transmission from the right probability amplitudes, which can be obtained from the solutions of the stationary Schr\"{o}dinger equation~\cite{Landau1977,Sakurai1993}.
Using expression \eqref{expresionS} and taking the partial trace over momentum, we write Eq.~\eqref{scattmap} in the eigenbasis of $H_Y$ as \cite{Barra2021}
    \begin{align}
    \label{scattmap2}
    \rho'_{j'k'} = \sum_{jk} \mathbb{S}_{j'k'}^{jk} (\rho_A \otimes \rho_B)_{jk} \; ,
    \end{align}
where $\rho'_{j'k'}\equiv \braket{j'|\rho'|k'}$ and 
\begin{align}
    \label{scatteringmap}
     \mathbb{S}_{j'k'}^{jk} & = \sum_{\alpha'=\pm}\int_{ p_{\rm inf}}^\infty dp\, \rho_X(p,\pi(p)) \sqrt{\frac{p}{\pi(p)}} \nonumber \\
     & \times s_{j'j}^{(\alpha' +)}(E_p+e_j)\left[s_{k'k}^{(\alpha' +)}(E_{p}-\Delta_{j'j}+e_{k'})\right]^* \; .
\end{align}

In the last expression, $\pi(p)=\sqrt{p^{2}-2m(\Delta_{j'j}-\Delta_{k'k})}$, and the  lower integration limit 
$ p_{\rm inf}$ is obtained from $ p^2_{\rm inf}/2m={\rm max}\{0,\Delta_{j'j},\Delta_{j'j}-\Delta_{k'k}\}$, which guarantees that the channels are open in the integration domain. As discussed in Ref.~\cite{Barra2021}, the scattering map shown in Eq.~\eqref{scattmap} or Eq.~\eqref{scattmap2} is completely positive and trace-preserving and does not generally lead to unitary dynamics. However, as we show next, it does so in the regime considered in this study.

\subsubsection{Scattering matrix}
\label{smsubsec}
The scattering matrix simplifies under conditions 1 and 2.
Indeed, as we show in appendix~\ref{SCR}, by solving the stationary Schr\"odinger equation under these conditions, we obtain that
reflection is negligible and the effect of the collision is a shift in the transmitted wave. Specifically, we show in appendix~\ref{SCR3} that the scattering matrix simplifies to
	\begin{align}
	s_{j'j}^{(\alpha' +)}(E_p) = &~ \delta_{\alpha'+}~\braket{j'|e^{-i \tau_pV/\hbar}|j} \label{eq12}\; ,
	\end{align}
with $\tau_p \equiv ma/p$ and $\tau_p V =(m/ p) \int_{-a/2}^{a/2} V(x)\nu dx$. In other words, we are justified in treating the potential $V(x)$ as an effective barrier of length $a$ and height $\langle V \rangle=(1/a)\int_{-a/2}^{a/2} V(x) dx$. A similar result has also been obtained for a potential barrier via transfer matrix methods in the semi-classical regime \cite{Parrondo2022}. 

\subsubsection{Particle's state of motion}

In Ref.~\cite{Parrondo2022} the semi-classical regime was used together with narrow states in momentum, which act as a heat source to the internal system when mixed with the effusion distribution. Here, we take the semi-classical regime with states which are broad in momentum and narrow in position, leading instead to internal unitary evolution. 

Under condition 3 (see appendix \ref{BWPA}), we have $\rho_X(p,\pi(p))\simeq \rho_X(p,p)$. 
Also, if $p_0^2\gg m\Delta_Y$ (condition 3) we have $s_{k'k}^{(\alpha' +)}(E_{p}-\Delta_{j'j}+e_{k'})\simeq s_{k'k}^{(\alpha' +)}(E_{p})$ and $\sqrt{p/\pi(p)}\simeq 1$. Under condition 2, the lower integration limit in Eq.~\eqref{scatteringmap} can be extended to minus infinity and Eq.~\eqref{eq12} be derived. 

In this regime, Eq.~\eqref{scatteringmap} is greatly simplified. Using $\rho_A \otimes \rho_B = \sum_{ji} (\rho_A \otimes \rho_B)_{ji} \ket{j}\bra{i}$, Eq.~\eqref{scattmap2} can be written in a basis-independent fashion as
    \begin{align}
    \label{scatteringmapsimple}
    \rho' = \int_{-\infty}^{\infty} dp~\rho_X(p,p) ~ e^{-i \tau_p V/\hbar}~ (\rho_A \otimes \rho_B)~e^{i \tau_p V/\hbar} \; ,
    \end{align}
which is a completely positive and trace preserving random unitary map~\cite{Audenaert2008}.

The last step to arrive at Eq.~\eqref{result} involves performing a saddle point approximation around $p_0$ to perform the integral in Eq.~\eqref{scatteringmapsimple}, which is possible since $p_0 \gg \sigma_p$ is fulfilled (see appendix.~\ref{SCR4}). We thus obtain our first result in Eq.~\eqref{result} and the second result follows immediately. We note that conditions 1-3 are sufficient conditions to obtain Eq.~\eqref{scatteringmapsimple} from Eq.~\eqref{scatteringmap} and imply that the entanglement between the joint internal degrees of freedom of $A$ and $B$ with the kinetic degree of freedom $X$ is negligible.

\subsection{Time-dependent model}

We now discuss how to derive the aforementioned results for the time-dependent model. The unitary operator $U_I(\tau)$ in Eq.~\eqref{tdmap} is the solution to the von Neumann equation 
    \begin{align}
    \label{vonneumann}
    \frac{d}{dt} U_I(t) = -\frac{i}{\hbar}V_I(t) U_I(t)
    \end{align}
where $V_I(t) = e^{i H_Y t /\hbar} V(t) e^{-i H_Y t /\hbar}$ is the interaction in the interaction picture. The solution of Eq.~\eqref{vonneumann} can be generally written as $U_I(\tau)=\exp{\Omega(\tau)}$, where $\Omega(\tau) =  \sum_{k=1}^{\infty} \Omega_k(\tau)$ is the Magnus expansion \cite{Blanes2009}, whose first two terms read
    \begin{align}
    \Omega_1(\tau) & = - \frac{i}{\hbar} \int_{-\tau/2}^{\tau/2} dt V_I(t) \; ,\label{s1} \\
    \Omega_2(\tau) & = - \frac{1}{2 \hbar^2} \int_{-\tau/2}^{\tau/2} dt  \int_{-\tau/2}^{t} dt' [V_I(t),V_I(t')] \; .
    \end{align}
The higher-order terms consist of linear combinations of nested commutators of $[V_I(t),V_I(t')]$. For instance, $ \Omega_3(\tau)$ contains integrals of terms such as $[V_I(t),[V_I(t'),V_I(t'')]]$ and so on. When the interaction time is very short compared to the internal dynamics (condition 1) we have $V_I(t) \simeq V(t)$. In this case,  $[V_I(t),V_I(t')] \simeq \tilde{V}(t)\tilde{V}(t')[\nu,\nu]=0$, due to the factorized form of the interaction. The evolution operator is determined by the first order term \eqref{s1} of the expansion, with $V_I(t) \simeq V(t)$ i.e., $U_I(\tau) = \exp{(-i \tau V/ \hbar)}$, using the definition in Eq.~\eqref{avVt} for $\langle \tilde{V} \rangle$. We thus conclude that Eq.~\eqref{tdmap} reduces to Eq.~\eqref{result} under condition 1.

\section{Applications} 
\label{secV}

\subsection{Collision of two spins}

To illustrate our results, we consider a numerical model where $A$ and $B$ are both 1/2-spins with Hamiltonians $H_{A} = \Delta_{A} \sigma^{z}_{A}$ where $2\Delta_{A}$ is the energy gap of $A$ and $\sigma^{i}_{A}$ are Pauli matrices $i=x,y,z$ in the Hilbert space of $A$ (equivalently for $B$). The internal interaction between the spins is given by $\nu = J_x~\sigma^{x}_A \otimes \sigma^{x}_B + J_y~\sigma^{y}_A \otimes \sigma^{y}_B$, where $J_x,J_y \in \mathbb{R}$. 

For the scattering map, we take a sinusoidal potential vanishing at the boundaries $V(x) =(\pi /2)  V_0\cos(\pi x/a)$ for $|x|<a/2$ and zero otherwise. The minimal length scale characterizing this potential is $a_{\rm min}\sim a$. The exact scattering matrix $s_{j'j}^{(\alpha'\alpha)}(E)$ is computed by solving the non-linear equations of multi-channel scattering theory \cite{Razavy2003} summarized in appendix~ \ref{razavy-sec}. In this first part of this section, we consider the particle's state of motion to be a pure state $\rho_X=\ket{\phi}\bra{\phi}$, so that $\rho_X(p,\pi(p))=\phi(p)\phi^*(\pi(p))$ with $\phi(p) \equiv \braket{p|\phi}$; mixed states are analyzed at the end of the section. We thus consider a Gaussian state
\begin{align}
\label{gaussianpurestate}
\phi(p) = \frac{ \exp[-(p-p_0)^2 / 4 \sigma_p^2 - ipx_0/\hbar]}{(2 \pi \sigma_p^2)^{1/4}}
\end{align}
with average and variance in momentum given respectively by $p_0$ and $\sigma_p^2$, while the average position and variance in position is $x_0$
and $\sigma_x^2 = \hbar^2/(4\sigma_p^2)$. The state is normalized according to $\int dp \braket{p|\rho_X|p} = \int dp |\phi(p)|^2 = 1$. All this information is plugged into Eq.~\eqref{scatteringmap} which in turn is used in the scattering map in Eq.~\eqref{scattmap2}. 

Regarding the time-dependent model, we choose a triangular function as potential: $\tilde{V}(t) = (4/\tau) V_0 (\tau/2-|t|)$ for $|t|<\tau/2$ and zero otherwise with $V_0 > 0$.  The exact dynamics is computed by solving Eqs.~\eqref{vonneumann} and \eqref{tdmap}. Note that, as required by our theory, the time-dependent and space-dependent potentials satisfy $\langle \tilde{V}\rangle=\langle V\rangle=V_0$ and the interaction times are such that $\tau = \tau_{p_0}$.

Regarding our result in Eq.~\eqref{result}, the unitary transformation of our analytical model  in the eigenbasis of $H_Y$ ($\ket{\uparrow\uparrow},\ket{\uparrow\downarrow},\ket{\downarrow\uparrow},\ket{\downarrow\downarrow}$) is the $4 \times 4$ matrix
\begin{widetext}
\[
e^{-i\lambda\nu}=
\left(
\begin{array}{cccc}
\cos \lambda_1  &0&0&-i\sin \lambda_1 \\
0&\cos \lambda_2   &-i\sin \lambda_2 &0\\
0&-i\sin \lambda_2  &\cos \lambda_2 &0\\
-i\sin \lambda_1 &0&0&\cos \lambda_1 
\end{array}
\right)
\]
\end{widetext}
where we define the dimensionless parameter $\lambda \equiv V_0\tau/\hbar$ quantifying the interaction strength, $\lambda_1 \equiv \lambda (J_x - J_y)$ and $\lambda_2 \equiv \lambda (J_x + J_y)$. The last matrix can then be reordered and written as 
\begin{align}
e^{-i\lambda\nu}= e^{-i\lambda_1 \sigma^x_1}\oplus e^{-i\lambda_2 \sigma^x_2}
\label{eilambda}    
\end{align}
a direct sum on two two-dimensional subspaces
$\{\ket{+}_1\equiv\ket{\uparrow\uparrow}, \ket{-}_1\equiv\ket{\downarrow\downarrow}\}$ and $\{\ket{+}_2\equiv\ket{\uparrow\downarrow}, \ket{-}_2\equiv\ket{\downarrow\uparrow}\}$. Similarly, with the same order for the basis, the Hamiltonian 
\begin{equation}
\label{Hspin}
H_Y=\{(\Delta_A+\Delta_B)\sigma^z_1\}\oplus\{ (\Delta_A-\Delta_B)\sigma^z_2\}
\end{equation}
is a direct sum. In Eqs.~\eqref{eilambda} and \eqref{Hspin}, $\sigma^i_1$ are the Pauli matrix in the basis $\{\ket{+}_1,\ket{-}_1\}$ and similarly for $\sigma^i_2.$ In summary, the dynamics given by Eq.~\eqref{result} is  here equivalent to the oscillatory dynamics of two independent spins $1$ and $2$, oscillating with period $\pi$ within two independent sectors $\{\ket{\uparrow\uparrow},\ket{\downarrow\downarrow}\}$ and
$\{\ket{\downarrow\uparrow},\ket{\uparrow\downarrow}\}$, respectively. In terms of $\lambda$, spin $1$ completes $n$ cycles at $\lambda = \pi n/(J_x - J_y)$ and spin $2$ at $\lambda = \pi n /(J_x + J_y)$, with $n = 0,1,2...$. 

\begin{figure*}[t!]
\centering
\begin{subfigure}
  \centering
  \includegraphics[width=7.75cm]{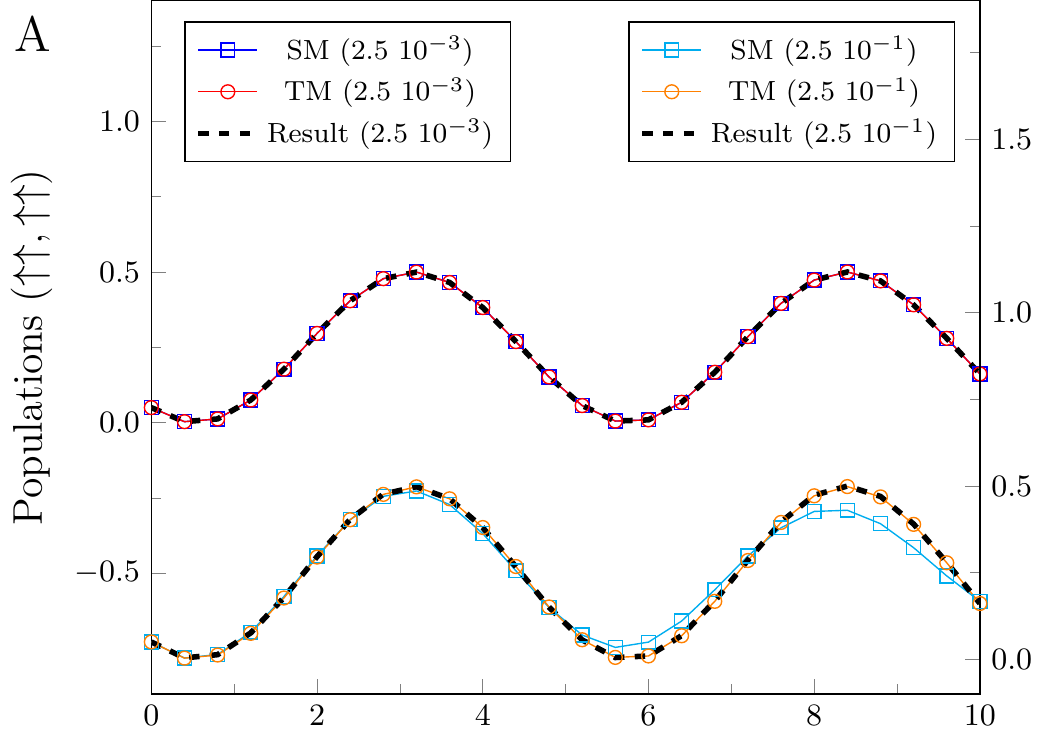}
\end{subfigure}
\begin{subfigure}
  \centering
  \includegraphics[width=7.9cm]{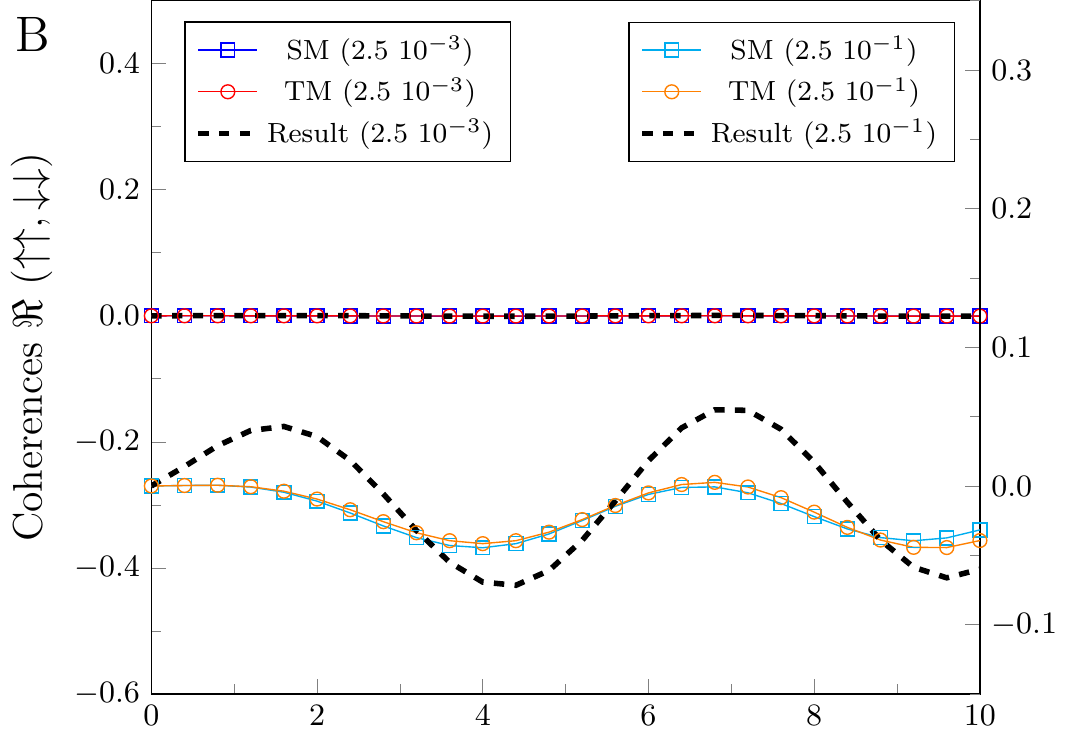}
\end{subfigure}
\begin{subfigure}
  \centering
  \includegraphics[width=7.75cm]{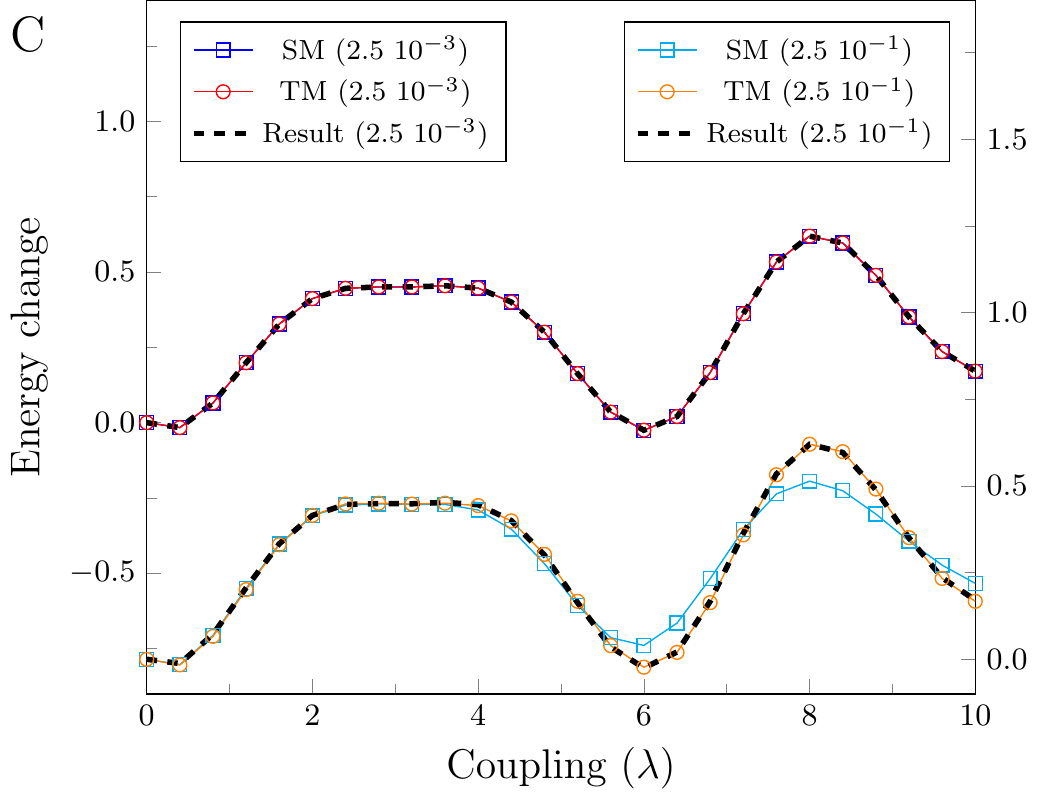}
\end{subfigure}
\begin{subfigure}
  \centering
  \includegraphics[width=7.75cm]{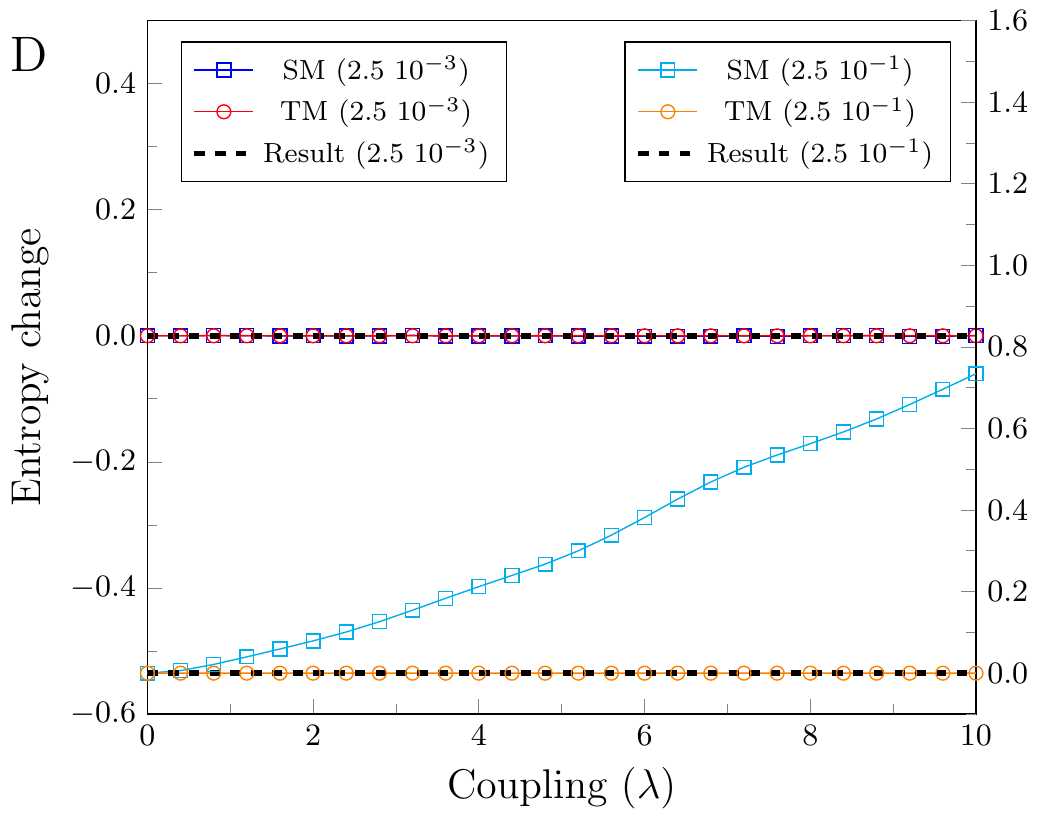}
\end{subfigure}
\caption{\label{fig:results} Upper panels: Populations (panel A) and real part of coherences (panel B) of a two spin-1/2 system after one collision according to the exact scattering map (label SM) in Eq.~\eqref{scattmap} (squares), time-dependent model (label TM) in Eq.~\eqref{tdmap} (circles) and our result in Eq.~\eqref{result} (black dashed lines) as a function of coupling parameter $\lambda \equiv V_0 \tau / \hbar$. We vary $\lambda$ by varying $V_0$ while keeping $\tau$ fixed at $\tau = 2.5 \times 10^{-3}$ (plots with dark color markers, read on the left axis) or $\tau = 2.5 \times 10^{-1}$ (plots with light markers, read on the right axis). Lower panel: Equivalent of upper panels for energy (panel C) and entropy changes (panel D), respectively. 
The initial state of the both spins is pure: in the eigenbasis of $H_Y$, the state of $A$ is $(\rho_{A})_{\uparrow \uparrow}=0.1$, $(\rho_{A})_{\downarrow \downarrow} = 1 - (\rho_{A})_{\uparrow \uparrow}$ and $(\rho_{A})_{\uparrow \downarrow} = \sqrt{|(\rho_{A})_{\uparrow \uparrow} (\rho_{A})_{\downarrow \downarrow}|} \exp(i \pi /4)$, similarly for $B$ with $(\rho_{B})_{\uparrow \uparrow}=0.5$ instead. 
The model parameters are $\Delta_A = 3/4$ and $\Delta_B = 1/2$ (non-degenerate spins), $J_x=0.8$, $J_y=0.2$, $\hbar=m=1$, $a=3.5$. Note that $\tau_{p_0}=\tau\Rightarrow p_0=ma/\tau$ and $\sigma_p \gg 20 m \Delta_Y / p_0$ for scattering map.}
\end{figure*}

\subsection{Numerical results}

\subsubsection{Pure states}

In Fig.~\ref{fig:results}, we display the state of $A$ and $B$ in the first sector $\{\ket{\uparrow\uparrow},\ket{\downarrow\downarrow}\}$ (populations $\bra{\uparrow\uparrow}\rho'\ket{\uparrow\uparrow}$ in panel A and coherences $\bra{\downarrow\downarrow}\rho'\ket{\uparrow\uparrow}$ in panel B), as well as the energy and entropy changes after the interaction (panels C and D), as a function of the coupling parameter $\lambda \equiv V_0 \tau / \hbar$. We increase $\lambda$ by increasing $V_0$ while keeping $\tau$ fixed, and since we require $\tau = \tau_{p_0}$, we take $p_0= ma/\tau$ in the scattering map. We display the results for $\tau = 2.5 \times 10^{-3}$ and $\tau = 2.5 \times 10^{-1}$, with condition 1 holding in the former case but not in the latter. Condition 2 and 3 are here always fulfilled. 

For $\tau= 2.5 \times 10^{-3}$ (high $E_{p_0}$), we observe a very good matching between the time-dependent model, the scattering map and our result in Eq.~\eqref{result}, even when the coupling is strong. Indeed, when $\lambda = 10$ we still have $E_{p_0} / V_0 \gg 1$ and thus all conditions for our result to hold are fulfilled. The pure state therefore excites both spins without changing their entropy with Rabi-like oscillations~\cite{Englert1991,Haroche1991,Haroche2006} of period $\lambda = 5 \pi /3 \simeq 5.24$ as expected ($J_x = 0.8$ and $J_y = 0.2$). For $\tau = 2.5 \times 10^{-1}$ (low $E_{p_0}$), we see that our result immediately departs from the exact scattering and time-dependent models at the level of coherences and entropy change. Remarkably, our result still replicates the populations and energy changes of scattering and time-dependent models provided the coupling is not too large (i.e., $E_{p_0} / V_0 \gg 1$ still holds), but they mismatch for larger couplings (i.e., $E_{p_0} / V_0 \sim 1$ and reflection is no longer negligible).

We tested many other potentials $\tilde{V}(t)$ and $V(x)$ with $\langle \tilde{V} \rangle=\langle V\rangle$ and confirmed  numerically that they induce the same dynamics in the two spins, provided the conditions for our theory hold.


\begin{figure*}[t!]
\centering
\begin{subfigure}
  \centering
  \includegraphics[width=7.75cm]{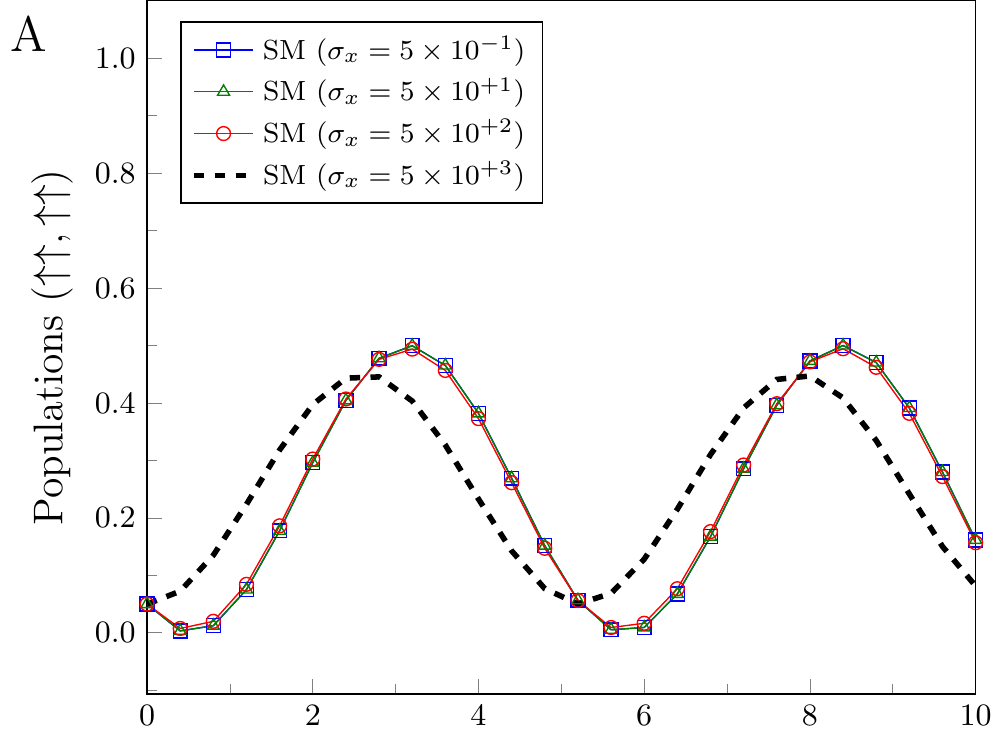}
\end{subfigure}
\begin{subfigure}
  \centering
  \includegraphics[width=7.75cm]{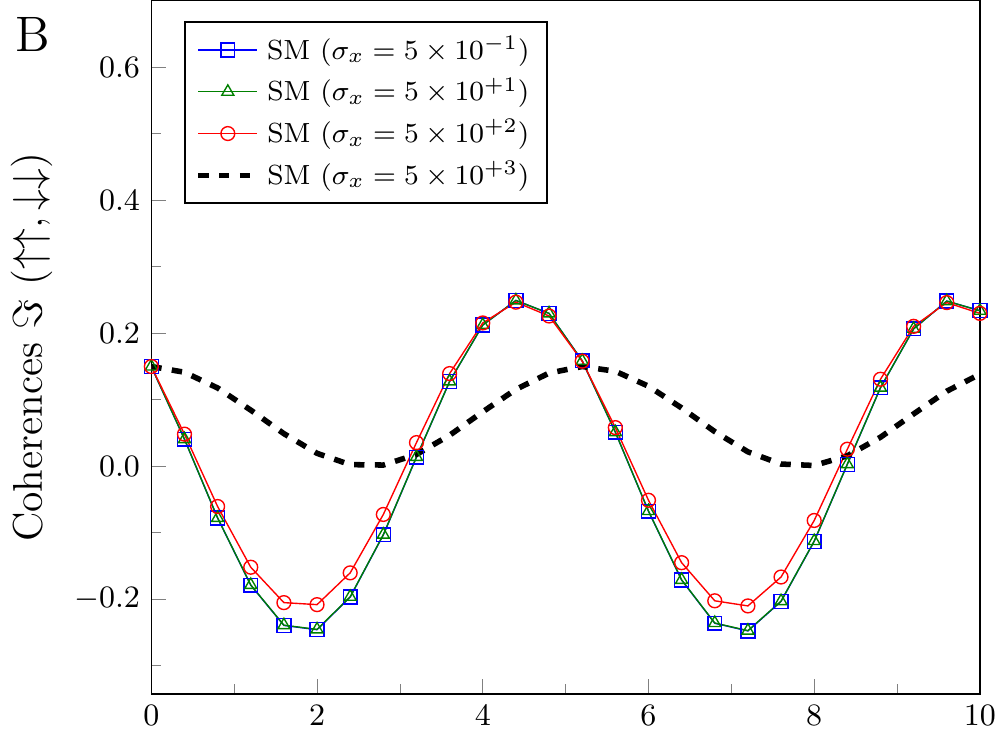}
\end{subfigure}
\begin{subfigure}
  \centering
  \includegraphics[width=7.75cm]{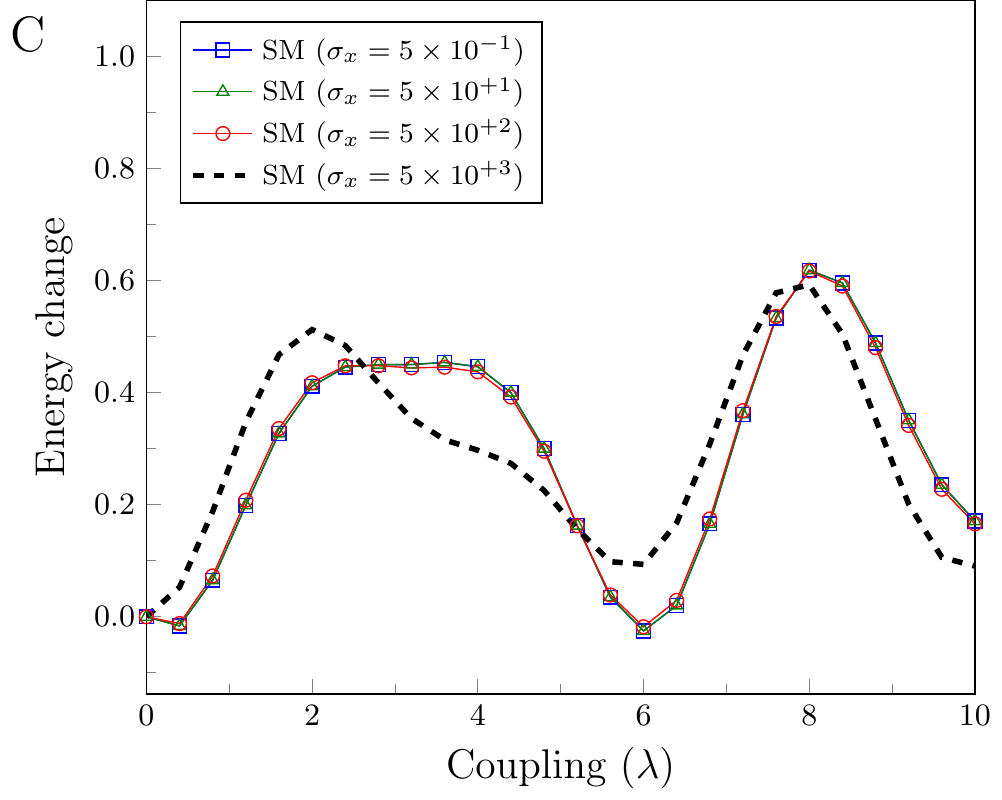}
\end{subfigure}
\begin{subfigure}
  \centering
  \includegraphics[width=7.75cm]{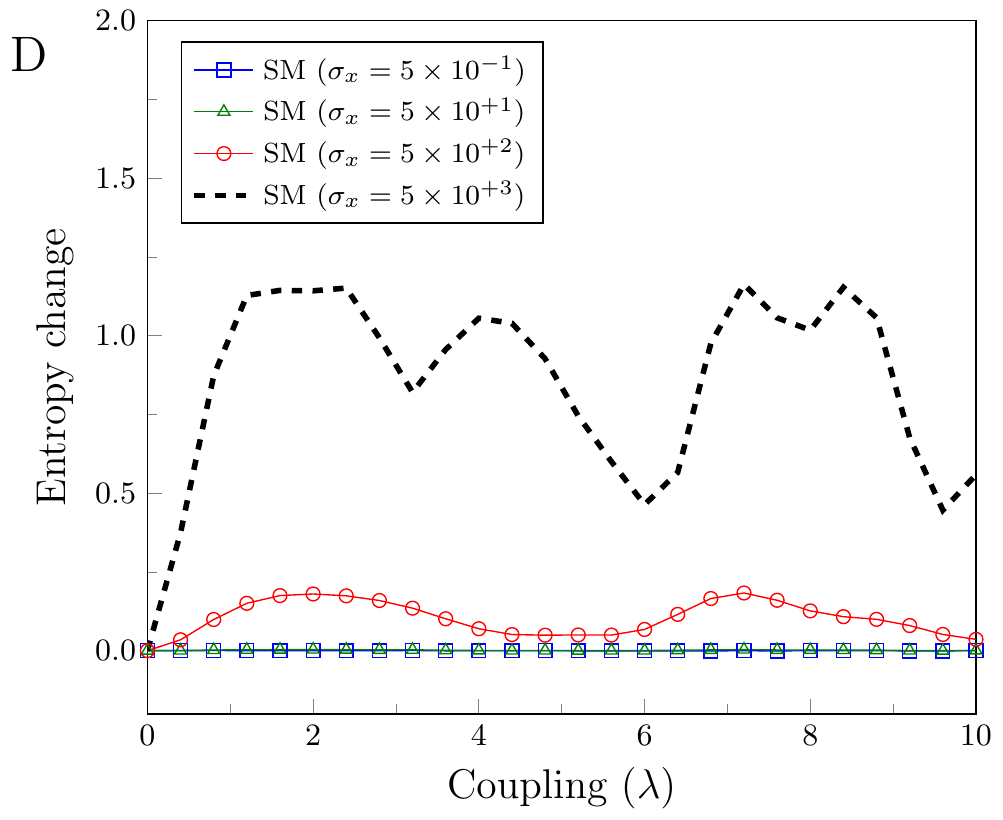}
\end{subfigure}
\caption{\label{fig:results-mixed} Populations (panel A), imaginary part of coherences (panel B), energy and entropy changes (panels C and D) of a two spin 1/2 system after one collision according to the exact scattering map (label SM) in Eq.~\eqref{scattmap}, as a function of coupling parameter $\lambda \equiv V_0 \tau / \hbar$. We show the results for fixed $\tau = 2.5 \times 10^{-1}$ and different $\sigma_x$ thus changing the purity of the states of motion $P=\hbar /(2 \sigma_p \sigma_x)$. For $\sigma_x = 0.5$ we have $P=1$ so the state is pure (squares). For larger values $\sigma_x = 50,500,5000$ we have $P=10^{-2},10^{-3},10^{-4}$ so the state is mixed. The remaining parameters are the same as in Fig.~\ref{fig:results}.}
\end{figure*}

\subsubsection{Mixed states
}
Now we consider Gaussian mixed states with Wigner functions
\begin{align}
 W(p,x)=\frac{\exp[-(p-p_0)^2/2\sigma_p^2-(x-x_0)^2/2\sigma_x^2]}{2\pi\sigma_p\sigma_x} \; .
\end{align}
As discussed in appendix \ref{BWPA}, these states generalize the pure state considered before, i.e. the position and momentum are Gaussian with averages $x_0$ and $p_0$ and variances $\sigma_p^2$ and $\sigma_x^2$, respectively. When $\sigma_x=\hbar/(2\sigma_p)$, the state is pure $\rho_X(p,p') = \phi(p)\phi^*(p')$ with $\phi(p)$ given by Eq.~\eqref{gaussianpurestate}, but when $\sigma_x > \hbar/(2\sigma_p)$ the state is mixed (see Eq.~\eqref{gaussianmixedstate}). We predict that, as long as the state is narrow in position with respect to the system, i.e. $\hbar/\sigma_x \gg m\Delta_Y/p_0$ in condition 3, the scattering map should still be unitary, provided that conditions 1-2 hold.

In Fig.~\ref{fig:results-mixed}, we start from the minimum value $\sigma_x = \hbar/(2\sigma_p) = 0.5$ and  increase $\sigma_x$ while fixing the value of $m\Delta_Y/p_0$. For $\sigma_x=0.5$, the state is pure and the data correspond to Fig.~\ref{fig:results}, where all conditions, including the aforementioned inequality, are satisfied. When $\sigma_x=50$, the quantities $m\Delta_Y/p_0$ and  $\hbar/\sigma_x$ are of the same order, but still the scattering map behaves unitarily. For larger values $\sigma_x \geq 500$, the aforementioned inequality is violated and we observe significant changes in the dynamics, which now induces energy and entropy exchanges. The high value of entropy change signals the breakdown of unitary evolution, so the mixed state no longer acts as a work source for the internal system.

\section{Conclusions} 
\label{secVI}

We have considered the effect of a collision between a fixed system and a fast particle described by a state of motion squeezed in position and broad in momentum, showing that the map describing the effect of the collision on the joint internal degrees of freedom becomes unitary. 

The possibility of eliminating kinetic degrees of freedom in favor of a time-dependent description, valid in the semi-classical limit, has been discussed before 
\cite{Aharonov1961}. Here, we extended such analysis to the case of particles with internal structure. By doing this, we deduced the semi-classical expression for the scattering matrix which we used to prove our results. Together with an analysis of the quantum state of motion of the particle, this lead us to conditions 1-3 which define the regime where the scattering map behaves unitarily according to Eq.~\eqref{result}. Since the energy transfers within the internal system occur without entropy change, such collisions can be used to model the effect of a work source.

This finding nicely complements the results of Refs.~\cite{Barra2021,Parrondo2022} where we showed that collisions with effusing mixtures of incoming narrow packets can model heat sources. 
The width of the packet is crucial to discriminate between these two cases. 
In the narrow case, the states of motion associated with each transition in the scattering map are distinguishable, so the particle's state of motion carries away information about the transitions in the internal system, acting as a measurement apparatus. Instead, very fast and broad states are indistinguishable and they do not reveal information about the internal transitions, resulting in coherent evolution of the internal system.

Most collisions do not behave either as heat sources or work sources, but can probably be analyzed as generic free energy sources. Pure entropy collisions may also occur. This opens the way for many further explorations that may eventually lead to a scattering-based formulation of quantum thermodynamics.  


\acknowledgments{
SLJ is supported by the Doctoral Training Unit on Materials for Sensing and Energy Harvesting (MASSENA) with the grant: FNR PRIDE/15/10935404.
ME is also funded by the European Research Council (project NanoThermo, ERC-2015-CoG Agreement No. 681456) and the Foundational Questions Institute Fund (Grant number FQXi-IAF19-05). 
F. B. thanks Fondecyt project 1191441 and ANID – Millennium Science Initiative Program – NCN19\_170.
Part of this work was conducted at the KITP, a facility supported by the US National Science Foundation under Grant No. NSF PHY-1748958. JMRP acknowledges financial support from the Spanish Government (Grant FLUID, PID2020-113455GB-I00) and from the Foundational Questions Institute Fund, a donor advised fund of Silicon Valley Community Foundation (Grant number FQXi-IAF19-01).
}

\appendix

\section{Semi-classical regime}
\label{SCR}

\subsection{Introduction}

Consider the time-independent Schr\"{o}dinger equation in the absence of internal degrees of freedom, with the potential $V(x)$ being effective over the region $x \in (-a/2,a/2)$. If we take a position very far away to the left of this region, then $ \psi(x) \sim \exp{(\mathrm{i}px/\hbar)}$ is a solution to the equation, representing a plane wave associated to a particle which travels free from the potential. In general, inside the region of the potential $V(x)$ the solutions are not plane waves. However, it is well known that if the particle is fast $E_p \gg V(x)$ and if its de Broglie wave length is much shorter than the scales over which the potential varies significantly $p a_{min} \gg \hbar$, where $a_{min}$ is this scale, then the effect of the potential on the wave can be simplified. The plane wave at position $x$ inside the potential is then multiplied by a phase proportional to the integral of the interaction \cite{Landau1977,Sakurai1993}. More precisely, we have
    \begin{align}
    \label{phaseshift-ap}
    \psi(x) \sim \exp{\Big(\frac{\mathrm{i}px}{\hbar}\Big)} \exp{\Big(-\frac{\mathrm{i}m}{\hbar p} \int_{-\infty}^{x}V(x') ~dx' \Big)} \; .
    \end{align}
By taking $x \rightarrow +\infty$ the the total phase shift due scattering with the potential, which is proportional to the scattering amplitude is recovered \cite{Landau1977}. Since the potential is supported on the interval $x \in (-a/2,a/2)$, we have $ \int_{-a/2}^{a/2} V(x) dx = \langle V \rangle a$. In other words, the potential can be effectively treated as a barrier of length $a$ and height $\langle V \rangle$. The purpose of this section is to show that the same is true in the presence of internal degrees of freedom, which then allows us to simplify the scattering matrix appearing in Eq.~\eqref{scatteringmap}.

\subsection{Derivation}

Let $\ket{\psi}$ be a solution to the time-independent Schr\"{o}dinger equation 
\begin{equation}
H\ket{\psi} =\left[\frac{p^2}{2m}+H_Y+\mathcal{V}(x)\right]\ket{\psi}= E \ket{\psi}
\label{scatteringhamiltonian-ap}
\end{equation}
with some energy $E$ and $\ket{\psi_0}$ the corresponding free solution with the same energy, valid very far away from the potential. Projecting Eq.~\eqref{scatteringhamiltonian-ap}, onto the position eigenbasis $\braket{x|H|\psi} = E \braket{x|\psi}$, we obtain an operator equation for $\ket{\psi(x)} \equiv \braket{x|\psi}$ in the Hilbert space of $Y$
    \begin{align}
    \label{schrodinger-ap}
    \Big( \hbar^2\frac{ d^2}{dx^2} + \mathcal{P}^2 \Big) \ket{\psi(x)} = 2m \mathcal{V}(x) \ket{\psi(x)} \; ,
    \end{align}
where $d/dx$ is a total derivative and the interaction is $\mathcal{V}(x) \equiv V(x) \nu$. The momentum operator is defined as
    \begin{align}
    \label{operadormomentum-ap}
    \mathcal{P} \equiv \sqrt{2m (E-H_Y)} \; 
    \end{align}
and we assume that $E$ is larger than the maximum eigenvalue of $H_Y$, in which case $\mathcal{P}$ is a positive operator and thus self-adjoint. To make progress in solving Eq.~\eqref{schrodinger-ap}, we look for solutions of the form
    \begin{align}
    \label{solutions-ap}
    \ket{\psi(x)} = e^{i \mathcal{P}x/\hbar} \ket{\Psi(x)} \; ,
    \end{align}
where the exponential operator is unitary. 
Substituting in Eq.~\eqref{schrodinger-ap} and noting that $[\exp(i \mathcal{P}x/\hbar),\mathcal{P}]=0$, we verify that $\ket{\Psi(x)}$ satisfies
    \begin{align}
    \label{schrodinger2-ap}
    \Big( \frac{\hbar^2}{2} \mathcal{P}^{-1} \frac{d^2}{dx^2} + i \hbar \frac{d}{dx} \Big)\ket{\Psi(x)} = \mathcal{V}_{\mathcal{P}}(x) \ket{\Psi(x)} \; ,
    \end{align}
where the operator $\mathcal{V}_{\mathcal{P}}(x')\equiv e^{-i \mathcal{P}x'/\hbar} m \mathcal{P}^{-1}\mathcal{V}(x') e^{i \mathcal{P}x'/\hbar}$ has units of momentum. The last expression is completely equivalent to Eq.~\eqref{schrodinger-ap}. Since we are interested in taking the semi-classical limit where $\hbar$ is very small compared to some action, we ignore the second derivative in the equation above. After we obtain the solution, we derive exactly the conditions under which this is valid. We thus get the equation
    \begin{align}
   i\hbar \frac{d}{dx} \ket{\Psi(x)} = \mathcal{V}_{\mathcal{P}}(x)  \ket{\Psi(x)} \; ,
    \end{align}
which formally has the same form of a Schr\"odinger equation in the interaction picture,
where $x$ plays the role of time, $m\mathcal{P}^{-1}\mathcal{V}(x)$ is the interaction and $-\mathcal{P}$ the free Hamiltonian \footnote{Note that it is also possible to define in $\mathcal{P}$ with a negative sign in Eq.~\protect \eqref{operadormomentum-ap}, and therefore one has two Schr{\"o}dinger equations, one for the ``forward'' time $x$ and another for the ``backward''. For our purposes we need only the former.}. This Schr\"{o}dinger equation is integrated with an ``initial'' position $x_0$ and an ``initial'' state with the same energy $E$ appearing in Eq.~\eqref{operadormomentum-ap}. We take the asymptotic state $\ket{\psi_0}$ introduced above, and through Eq.~\eqref{solutions-ap}, the corresponding $\ket{\Psi_0}$ to pick the ``initial'' condition. The evolution operator associated with this   
equation can be written in terms of a Magnus series as we did in section~\ref{sec:td}. In analogy to Eq.~\eqref{s1}, we have the first order term of the expansion
  \begin{align}
    \Omega_1(x) = -\frac{i}{\hbar} \int_{x_0}^{x} \mathcal{V}_{\mathcal{P}}(x')~dx'   \; .
    \end{align}
The higher order terms $\Omega_n(x)$ contain  $[\mathcal{V}_{\mathcal{P}}(x),\mathcal{V}_{\mathcal{P}}(x')]$ and a sequence of nested commutators of it. Now we show that one can neglect the higher order terms in the Magnus expansion when the kinetic energy is sufficiently large.  
For large $E$ we have
\begin{widetext}
   \begin{align}
    \mathcal{P} = \sqrt{2m (E -H_Y)} = \sqrt{2mE}  - \sqrt{\frac{m}{2E}}H_Y  \Big[1- \mathcal{O}\Big(\frac{H_Y}{E} \Big)^2 \Big] \; ,
    \end{align}
allowing the replacement 
\begin{align}
\mathcal{V}_{\mathcal{P}}(x') \simeq e^{ i\sqrt{\frac{m}{2E}} H_Yx'/\hbar}\frac{ m}{\sqrt{2mE}}\left[ 1+\frac{H_Y}{E}\right]\mathcal{V}(x') e^{- i \sqrt{\frac{m}{2E}}H_Yx'/\hbar} \simeq \sqrt{\frac{m}{2E}}\mathcal{V}(x')
\end{align}
\end{widetext}
where in the last approximation we used that $|x'|<a/2$ and considered $\sqrt{2E/m}\gg a\Delta_Y/\hbar$, eliminating the exponentials of the expression at the left. In this limit we have $[\mathcal{V}_{\mathcal{P}}(x),\mathcal{V}_{\mathcal{P}}(x')]=m(V(x)V(x')/2E)[\nu,\nu]=0$ and all higher orders can be neglected. The above inequality $\sqrt{2E/m}\gg a\Delta_Y/\hbar$ allows us to consider $E \simeq E_p$, which is equivalent to $ \tau_p\Delta_Y/\hbar\ll 1$
(see condition 1 of the main text). Thus, the Magnus series reads
    \begin{align}
    \Omega(x) & = -\frac{i}{\hbar}\sqrt{\frac{m}{2E}} \int_{x_0}^{x} \mathcal{V}(x')~dx' \nonumber \\ 
    & = -\frac{i m}{\hbar p} \int_{x_0}^{x} \mathcal{V}(x')~dx' \; ,
    \end{align}
meaning that
    \begin{align}
    \label{phaseshiftint-ap}
    \ket{\Psi(x)} = \exp{\Big( -\frac{im}{\hbar p} \int_{x_0}^{x} V(x') \nu dx' \Big)}\ket{\Psi_0(x_0)} \; .
    \end{align}
The last expression applied to Eq.~\eqref{solutions-ap} is the generalization of Eq.~\eqref{phaseshift-ap} in the presence of internal degrees of freedom. Now that we have a closed expression, we can verify the conditions under for the term proportional to $\hbar^2$ in Eq.~\eqref{schrodinger2-ap} to be negligible. Differentiating Eq.~\eqref{phaseshiftint-ap} twice with respect to $x$ and multiplying by $\hbar^2/p^2$, we obtain
    \begin{align}
    \frac{\hbar^2}{p^2} \frac{d^2}{dx^2} & \ket{\Psi(x)} = \nonumber \\ & -\Big(\frac{i m \hbar V'(x)}{p^3} + \frac{ m^2 V(x)^2}{p^4} \Big) \nu \ket{\Psi(x)} \; .
    \end{align}
For high momentum, the term proportional to $\hbar^2$ is negligible in comparison to the other terms in Eq.~\eqref{schrodinger2-ap} if the potential varies very slowly $p^3/(2m \hbar) \gg V'(x)$, a condition well known from semi-classical approximations in quantum mechanics \cite{Landau1977}. We simplify this condition by integrating over a minimum scale $a_{min}$ where the potential varies significantly by an amplitude $\Delta V$ obtaining $(E_p / \Delta V) p a_{min} / \hbar \gg 1$. Since we are interested in high kinetic energies, in the worst case we have $E_p / \Delta V \sim 1$ and thus $p a_{min} / \hbar \gg 1$ is a sufficient condition. 

\subsection{The scattering matrix}
\label{SCR3}

The expression for the scattering matrix can be straightforwardly obtained. In terms of the original wave function, Eq.~\eqref{phaseshiftint-ap} is
    \begin{align}
    \label{phaseshiftint2-ap}
    \ket{\psi(x)}  & =~ e^{ip_0x/\hbar} \exp{\Big(\frac{m}{i \hbar p} \int_{x_0}^{x} V(x') \nu dx' \Big)} \nonumber \\ & \times e^{-ip_0x_0/\hbar}\ket{\psi_0(x_0)} \; .
    \end{align}
From this expression we can deduce the transmission coefficient. Taking $x>a/2$ and $x_0<-a/2$, considering
$(m/p)\int_{x_0}^{x}V(x')dx'=\tau_p\langle V\rangle$, recalling the definition $V \equiv \langle V \rangle \nu$ and projecting Eq.~\eqref{phaseshiftint2-ap} on the left with $\bra{j'}$ we have
    \begin{align}
    \label{eqarriba}
\braket{j'|\psi(x)} =\braket{j'|e^{ip_0x/\hbar} e^{-i\tau_pV/\hbar}e^{-ip_0 x_0/\hbar}|\psi_0(x_0)} \; .
    \end{align}
Taking $\ket{\psi_0(x_0)}=e^{i p_j x_0/\hbar}\ket{j}$ with $p_j=\sqrt{2m(E-e_j)}$ in Eq.~\eqref{eqarriba} we obtain
    \begin{align}
 \braket{j'|\psi(x)} =e^{i p_{j'} x / \hbar}\braket{j'| e^{-i \tau_pV/\hbar}|j} = t_{j'j}e^{i p_{j'}x/\hbar} \; ,
    \end{align}
with $p_{j'}=\sqrt{2m(E-e_{j'})}$. The elements of the transmission matrix ${\bf t}$ are $t_{j'j}= \braket{j'| e^{- i\tau_pV / \hbar}|j}$. Since ${\bf t}$ is unitary, the reflection coefficients vanish in this limit. Thus, the scattering matrix under the conditions stated above is
	\begin{align}
	\label{scattansatz-ap}
	s_{j'j}^{(\alpha' +)}(E_p) = &~ \delta_{\alpha'+}~\braket{j'|e^{-i \tau_pV/\hbar}|j} \; ,
	\end{align}
as presented in the main text. It is valid when $\tau_p \Delta_Y / \hbar \ll 1$ (condition 1), $E_p \gg V(x)$ and $p a_{min} \gg \hbar$ (condition 2) and $E_p \gg \Delta_Y$ (inequality present in condition 3 of the main text).

\section{Mixed states}
\label{BWPA}

In this section, we want to generalize the notion of narrow and broad states of motion, presented in Ref.~\cite{Barra2021} for pure states, to mixed states. 
We also show that $\rho_X(p,\pi(p))\simeq \rho_X(p,p)$ for fast and broad mixed states of motion (condition 3).

\subsection{Wigner function}

To start, it is useful to consider the Wigner function, which is a quasi-probability distribution in classical phase space associated to a quantum state $\rho_X$
\begin{align}
    \label{wignerfunction}
    W(p,x)=\frac{1}{2 \pi\hbar}\int \rho_X(p+q/2,p-q/2)e^{i qx/\hbar} dq \; ,
\end{align}
where $\rho_X(p,p') \equiv \braket{p|\rho_X|p'}$ and the integral runs over all momentum space. Conversely, we can compute a quantum state $\rho_X$ starting from a given Wigner function, corresponding to the inverse of Eq.~\eqref{wignerfunction}
\begin{align}
    \label{quantumstate}
    \rho_X(p,p') = \int dx~W\Big(\frac{p+p'}{2},x\Big) e^{-i (p-p')x/\hbar} \; .
\end{align} 
It is useful to consider
\begin{align}
    \label{quantumstatewigner}
   \widetilde{W}(u,v) \equiv \int dx~W(u,x) e^{-i v x/\hbar} \;
\end{align} 
in terms of which we have
$\rho_X(p,p')=\widetilde{W}(u,v)$ with 
$u = (p+p')/2$ and $v = p-p'$. Importantly, if
$W(p,x) \in \mathbb{R}$ is supported in a region
around $(p_0,x_0)$ with characteristic width in
$x$ given by $\sigma_x$ and in $p$ by
$\sigma_p$, then $\widetilde{W}(u,v) \in
\mathbb{C}$ is supported in a region around
$(p_0,0)$ with characteristic width $\sigma_p$
in $u$ and at least $\hbar / (2\sigma_x)$ in
$v$, with the uncertainty relation $\sigma_p
\sigma_x \geq \hbar / 2$ holding for any
admissible quantum state. Furthermore, when
$p=p'$ we have $\widetilde{W}(p,0) = \rho_X(p,p)
\in \mathbb{R}$ which is a classical momentum
distribution, normalized over all momentum. 

Lastly, we can compute the purity of $\rho_X$ from the Wigner function as follows
\begin{align}
    \label{purity}
    P = \mathrm{Tr}[\rho_X^2] = 2 \pi \hbar \int dx \int  dp~W(p,x) \leq 1 \; ,
\end{align}
with the equality holding for pure states.

\subsection{Narrow and broad states}

We can now establish the narrow and broad wave packet distinction for mixed states. The crucial quantity to do this is $\rho_X(p,\pi(p))$ appearing in the scattering map of Eq.~\eqref{scatteringmap}, resulting from $\rho_X(p,p')$ after substituting $p'$ by $\pi(p) =\sqrt{p^{2}-2m(\Delta_{j'j}-\Delta_{k'k})}$. In other words, each element of the scattering map $\mathbb{S}_{j'k'}^{jk}$ is determined by the state $\rho_X(p,p')$ integrated along the line $p' = \pi(p)$. In terms of  Eq.~\eqref{quantumstatewigner}, the state is $\widetilde{W}(u,v)$ integrated along the hyperbola $u v = m (\Delta_{j'j}-\Delta_{k'k})$. Since the state is supported in $u$ in the region $u = p_0 \pm \sigma_p$, we substitute in the equation for the hyperbola to get
   \begin{align}
    v = \frac{m (\Delta_{j'j}-\Delta_{k'k})}{p_0(1 \pm \sigma_p / p_0)} \simeq \frac{m (\Delta_{j'j}-\Delta_{k'k})}{p_0} \; ,
    \end{align}
where we assume that $p_0 \gg \sigma_p$. We can thus distinguish between those states which are narrow in momentum $\hbar/ 2\sigma_x \leq v$ (hyperbola lies outside the support) or broad in momentum $\hbar / 2 \sigma_x > v$ (hyperbola lies inside the support). As studied in Ref.~\cite{Barra2021}, narrow states always lead to decoherence while broad ones generally preserve coherences.

Thus, for broad states in momentum we can approximate the integration line as $p'=\pi(p)\simeq p$. However, as we see in the following paragraph, a further condition is needed to have $\rho_X(p,\pi(p))\simeq \rho_X(p,p).$

\subsection{Fast and broad states}

For the very fast and broad states in momentum used in this study, we thus have the inequality $p_0 \gg \sigma_p \geq \hbar/2\sigma_x \gg m \Delta_Y / p_0$ (condition 3), where we used the uncertainty relation; the equality $\sigma_p = \hbar/2\sigma_x$ holds for pure Gaussian states as we confirm below. Although condition 3 is essential to preserve the coherence of the scattering process, it is not enough to achieve $\rho_X(p,\pi(p)) \simeq \rho_X(p,p)$, which we require to derive Eq.~\eqref{scatteringmapsimple}. This is because $\rho_X(p,\pi(p)) \in \mathbb{C}$ can still differ from $\rho_X(p,p) \in \mathbb{R}$ by a complex phase. To see this, we go back to Eq.~\eqref{quantumstatewigner} and note that if we translate the Wigner function in space to the origin $W(p,x) \rightarrow W(p,x-x_0)$ then we have $\widetilde{W}(u,v) \rightarrow \widetilde{W}(u,v)e^{ix_0v/\hbar}$ with real $\widetilde{W}(u,v)$ for a symmetric $W(p,x-x_0)$ with respect to the origin. Such a phase is negligible when $\hbar/x_0 \gg m\Delta_Y/p_0$, so in this case $\rho_X(p,\pi(p)) \simeq \rho_X(p,p)$. Note that the last inequality corresponds to condition 1 if $x_0$ is outside the scattering region $|x_0|>a$, which is already fulfilled in this study. 

\subsection{Mixed Gaussian states}

We now consider the following Wigner function 
\begin{align}
 W(p,x)=\frac{\exp[-(p-p_0)^2/2\sigma_p^2-(x-x_0)^2/2\sigma_x^2]}{2\pi\sigma_p\sigma_x}
\end{align}
describing a Gaussian probability distribution in phase space centered around $(x_0,p_0)$ and with variances $\sigma^2_x$ in position and $\sigma_p^2$ in momentum. The quantum state associated to this distribution is obtained by inserting the last expression into Eq.~\eqref{quantumstate}, yielding
\begin{align}
    \label{gaussianmixedstate}
    \rho_X(p,p') & = (2 \pi \sigma_p^2)^{-1/2} \exp \Big[ -\frac{((p+p')/2 - p_0 )^2}{2 \sigma_p^2} \Big] \nonumber \\
    & \exp \Big[ -\frac{(p-p')^2 \sigma_x^2}{2 \hbar^2} \Big] \exp \Big[ -\frac{i(p-p')x_0}{\hbar} \Big] \; ,
\end{align}
and its purity in Eq.~\eqref{purity} is given by $P = \hbar / (2 \sigma_p \sigma_x) \leq 1$. When $\sigma_x = \hbar / 2\sigma_p$, the state is pure $P = 1$ and $\rho_X(p,p') = \phi(p)\phi^*(p')$, where $\phi(p)$ is given by Eq.~\eqref{gaussianpurestate}. Conversely, when $\sigma_x \geq \hbar / 2\sigma_p$, the state is mixed $P \leq 1$ and $\rho_X(p,p')$ is generally given by Eq.~\eqref{gaussianmixedstate}. Thus, if we fix $\sigma_p$ and increase $\sigma_x$ starting from $\sigma_x=\hbar/(2\sigma_p)$, the purity starts at $P=1$ and decays as $\sim 1/\sigma_x$. In this way, we produced the numerical results displayed in Fig.~\ref{fig:results-mixed}. 

Note that the phase factor in Eq.~\eqref{gaussianmixedstate} doesn't contribute when $\hbar/x_0 \gg m\Delta_Y/p_0$, corresponding to condition 1 when $x_0$ is outside the scattering region $|x_0|>a$.

\section{Averaging the interaction}
\label{SCR4}

Starting from Eq.~\eqref{scatteringmapsimple},
    \begin{align}
    \label{scatteringmapsimple-ap}
    \rho' = \int_{-\infty}^{\infty} dp~\rho_X(p,p)~ e^{-i \tau_p V/\hbar}~ (\rho_A \otimes \rho_B)~e^{i \tau_p V/\hbar} \; ,
    \end{align}
where we extended the lower integration limit from $p_{\rm inf}$ to minus infinity since $\rho_X(p,p) = (2 \pi \sigma^2)^{-1/2} \exp[-(p-p_0)^2/2 \sigma^2]$ is supported at very high kinetic energies. Using the spectral decomposition of the interaction $V = \sum_{\alpha} V_{\alpha} \ket{\alpha}\bra{\alpha}$, we can simplify the integral by studying the function in the exponent
	\begin{align}
	F_{\alpha \beta}(p) \equiv -\frac{(p-p_0)^2}{2 \sigma^2} - \frac{ima(V_{\alpha}-V_{\beta})}{\hbar p} \; .
	\end{align}
Performing a saddle point approximation assuming $p_0 \gg \sigma_p$, a condition already fulfilled for the states considered in this study, we seek the extrema $F'_{\alpha \beta}(p)=0$
    \begin{align}
    \Big( \frac{p}{p_0}-1 \Big) \frac{p^2}{ \sigma_p^2} = \frac{i p_{\alpha \beta}}{p_0} \; ,
    \end{align}
where $p_{\alpha \beta} \equiv ma (V_{\alpha}-V_{\beta})/\hbar$ has units of momentum. Thus, if $p_0 \gg \sigma_p$ then $p = p_0$ is an approximate solution corresponding to a maximum (as can be confirmed by computing the second derivative). We expand $F_{\alpha \beta}(p)$ to second-order around $p_0$ and perform the integral, obtaining our final result
    \begin{align}
    \label{results-ap}
    \rho' = e^{-i \tau_p V/\hbar}~ (\rho_A \otimes \rho_B)~e^{i \tau_p V/\hbar} \; ,
    \end{align}
with the interaction time $\tau_{p_0} \equiv ma/p_0$.
\begin{widetext}
\section{Multi-channel scattering equations}
\label{razavy-sec}

We present the multi-channel scattering equations which allow us to compute numerically the exact scattering matrix presented in Sec.~\ref{secV}. For a particle coming from the left, we have the following relations
    \begin{align}
    \label{smatrixt2-ap}
    s_{j'j}^{(- +)}(E)=\sqrt{\frac{|p'|}{|p|}}r_{j'j}(E) \quad{\rm and} \quad s_{j'j}^{(+ +)}(E)=\sqrt{\frac{|p'|}{|p|}}t_{j'j}(E)
    \end{align}
where $p'$ and $p$ the final and initial momentum before and after the transition, while $r_{j'j}(E)$ and $t_{j'j}(E)$ are the reflection and transmission coefficients. The latter can be found by solving the coupled multi-channel scattering equations
	\begin{align}
	\label{razavy}
	\frac{dr_{j'j}(x)}{dx} & = \sum_{n,m} \frac{im V(x)}{\hbar p_n} \big[ \delta_{j'n} e^{i p_n x/\hbar} + r_{j'n}(x) e^{- i p_n x/\hbar} \big] \nu_{nm} \big[ \delta_{mj} e^{i p_m x/\hbar} + r_{mj}(x) e^{- i p_m x/\hbar} \big] \; , \\ \nonumber
	\frac{dt_{j'j}(x)}{dx} & = \sum_{n,m} \frac{imV(x)}{\hbar p_n} \big[ t_{j'n}(x) e^{-i p_n x/\hbar} \big] \nu_{nm} \big[ \delta_{mj} e^{i p_m x/\hbar} + e^{-i p_m x/\hbar} r_{mj}(x) \big] \; ,
	\end{align}
\end{widetext}
where we omitted the dependence on energy $E$ and $p_j = \sqrt{2m(E-e_j)}$ in the last expression. These are a set of non-linear, coupled differential equations in space. They were derived by Razavy in the context of quantum tunneling \cite{Razavy2003}. By using the boundary conditions $r_{ji}(\infty) = 0$, $t_{ji}(\infty) = \delta_{ji}$, $r_{ji}(-\infty) = r_{ji}$ and $t_{ji}(-\infty) = t_{ji}$ we recover the reflection and transmission coefficients defined which then completely determine scattering matrix.


\bibliographystyle{unsrtnat}
\bibliography{references}

\end{document}